\documentclass[a4,letters,usenatbib]{mn2e}%{mnras}
\usepackage{amsmath}
\usepackage{amssymb} 
\usepackage{multirow}
\usepackage{textcomp}
\usepackage{microtype}
\usepackage{subfigure}
\usepackage{xcolor,ulem}
\usepackage{graphicx}
\usepackage{float}
\usepackage[draft]{hyperref}
\usepackage{mathptmx}
\usepackage{tabularx}
\usepackage{caption}
\usepackage{soul}

\newcommand\numberthis{\addtocounter{equation}{1}\tag{\theequation}}

\newcommand{\email}[1]{\mbox{\href{mailto:#1}{#1}}}

\newcommand{\Mc}{\mathcal{M}}
\newcommand{\Msol}{M_{\odot}}
\newcommand{\Ayr}{A_{\mathrm{1yr}}}
\newcommand{\Mstar}{{\cal M}_{*}}

\def\ltsima{$\; \buildrel < \over \sim \;$}
\def\simlt{\lower.5ex\hbox{\ltsima}}
\def\gtsima{$\; \buildrel > \over \sim \;$}
\def\simgt{\lower.5ex\hbox{\gtsima}}
\def\apj{Astrophysical Journal}                % Astrophysical Journal
\def\apjl{Astrophysical Journal}             % Astrophysical Journal, Letters

\def\mnras{MNRAS}            % Monthly Notices of the RAS
\def\prd{Phys.~Rev.~D}       % Physical Review D
\def\araa{ARA\&A}             % Annual Review of Astron and Astrophys
\def\nat{Nature}              % Nature
\def\pasp{PASP}    % Astronomy and Astrophysics Reviews
\def\sovast{Soviet Ast.}

\title[Astrophysics from PTAs]{Astrophysical constraints on massive black hole binary evolution from Pulsar Timing Arrays}
\author[H. Middleton et al.]{
Hannah Middleton,$^1$\thanks{E-mail: \email{hannahm@star.sr.bham.ac.uk}}
Walter Del Pozzo,$^1$
Will M.\ Farr,$^1$
Alberto Sesana$^1$
\newauthor and Alberto Vecchio$^1$
\\
  $^1$School of Physics \& Astronomy, University of Birmingham, Birmingham, B15 2TT, UK}

\date{Accepted \dots Received \dots; in original form \dots}

\pubyear{2015}

\begin{document}
\label{firstpage}
\pagerange{\pageref{firstpage}--\pageref{lastpage}}
\maketitle

\begin{abstract}
  We consider the information that can be derived about massive 
  black-hole binary populations and their formation history
  solely from current and possible future pulsar timing array (PTA)
  results.  We use models of the stochastic gravitational-wave
  background from circular massive black hole binaries with chirp
  mass in the range $10^6 - 10^{11} M_\odot$ evolving solely due to
  radiation reaction. Our parameterised models for the black hole
  merger history make only weak assumptions about the properties of
  the black holes merging over cosmic time.  We show that current PTA
  results place an upper limit on the black hole merger density which 
  does not depend on the choice of a particular merger history model, 
  however they provide no information about the redshift or mass 
  distribution. We show that even in the case
  of a detection resulting from a factor of 10 increase in amplitude
  sensitivity, PTAs will only put weak constraints on the source
  merger density as a function of mass, and will not provide any
  additional information on the redshift distribution. Without
  additional assumptions or information from other observations, a
  detection {\it cannot} meaningfully bound the massive black hole
  merger rate above zero for any particular mass.

\end{abstract}


\section{Introduction}
\label{sec:Introduction}
Massive black holes (MBHs) reside at the centre of most galaxies ~\citep[see {\it e.g.}][and references therein]{2013ARA&A..51..511K}, and are believed to have a central role in their evolution ~\citep[see {\it e.g.}][and references therein for a recent review]{Volonteri:2012}.
Mapping the population of MBHs, studying their properties, demographics, and their connection to the broader formation of structure is one of the open problems of modern astrophysics. This is however difficult to tackle, due to the large range of scales and the wide variety of physical processes involved. The MBH evolutionary path remains a highly debated subject with many competing hypotheses still in play. Currently favoured hierarchical structure formation scenarios imply frequent galaxy mergers \citep{1978MNRAS.183..341W}. As a result MBH binaries (MBHBs) should be quite common in the Universe \citep{1980Natur.287..307B,2003ApJ...582..559V}. To-date there is no confirmed observed MBHB, although a number of candidates exist \citep[see e.g.][and references therein]{2012AdAst2012E...3D} and tantalising claims have been recently made \citep{GrahamEtAl2015_PossibleSMBHBinary,LiuEtAl_SMBHB_redshift2:2015,GrahamEtAl_CatalinaSurvey:2015}.

A means to survey MBHBs is through the observation of gravitational waves (GWs) that these systems generate as they inspiral towards their final merger. The accurate timing of an array of highly-stable millisecond pulsars -- a Pulsar Timing Array~\citep[PTA,][]{FosterBacker:1990} -- provides a direct observational means to probe the cosmic population of MBHBs on orbital timescales of order of several years. Astrophysical modelling suggests that the radiation emitted by an ensemble of MBHBs produces a GW stochastic background in the frequency range $\sim 10^{-9} - 10^{-7}$ Hz, where PTAs operate~\citep{SesanaVecchioColacino:2008,SesanaVecchioVolonteri:2009, RaviEtAl:2012, Sesana:2013}. Such a background affects the time of arrival of radio pulses in a characteristic fashion ~\citep{Sazhin:1978, Detweiler:1979,HellingsDowns:1983}, which can be used to discriminate the signal from a plethora of other undesired effects \citep{LentatiEtAl_EPTA:2015}.

Over the last decade pulsar timing has been used to put progressively tighter constraints on gravitational radiation in this frequency regime, ~\citep[see, e.g.][]{JenetEtAl:2006}. More recently the three international consortia consisting of the Parkes Pulsar Timing Array, PPTA~\citep{ShannonEtAl_PPTA:2015,ShannonEtAl_PPTAgwbg:2013}, NANOGrav~\citep{DemorestEtAl_NANOGRAVgwbg:2013,NANOGrav9year:2015} and the European Pulsar Timing Array, EPTA~\citep{LentatiEtAl_EPTA:2015,vanHaasterenEtAl_EPTAerratum:2012, vanHaasterenEtAl_EPTAgwbg:2011}, which in collaboration form the International Pulsar Timing Array~\citep{HobbsEtAl_IPTA:2010}, have used data from observations of unprecedented sensitivity to place constraints that are starting to probe astrophysically interesting regions of the parameter space \citep{Sesana:2013}.

In this paper, we consider a GW stochastic background produced by MBHBs in circular orbits losing energy and angular momentum purely through GW emission. We use an analytical merger rate model which makes minimal assumptions about the cosmological history of MBHB evolution and can capture the key characteristics of simulation results to investigate the astrophysical implications of current ~\citep{ShannonEtAl_PPTAgwbg:2013,LentatiEtAl_EPTA:2015,NANOGrav9year:2015}, and future plausible~\citep{SiemensEtAl:2013, RaviEtAl:2015} PTA results (either an upper-limit or a detection). Because our model is fully general---not committing to any particular cosmological MBHB merger history---we can identify and separate features of the merger history that are constrained by PTA data alone, from those that can only be constrained by adopting a particular merger history (as {\it e.g.} done in ~\cite{ShannonEtAl_PPTAgwbg:2013} and \cite{NANOGrav9year:2015})---in other words, by applying a particular cosmological prior.  Because our model is capable of reproducing the MBHB cosmic population found in cosmological simulations for certain choices of parameters, our results will be consistent with (but much broader than) those that would be obtained under a choice of specific classes of MBHB merger history models.

In Section \ref{sec:ModelMethod} we describe our method and the model used for the merger rate. In Section \ref{sec:Results} we present our results for several upper limits and for a possible future detection, and we discuss their implications for the population of MBHBs. We present our conclusions in Section \ref{sec:Conclusions}.

\section{Model and method}
\label{sec:ModelMethod}

\subsection{Astrophysical model}
\label{sec:model}
For the standard scenario of circular binaries driven by radiation reaction only, the characteristic strain of the GW stochastic background, $h$ at frequency $f$ is \citep{Phinney:2001}:
\begin{align*}
  h^2(f) = \frac{4G^{5/3}}{3\pi^{1/3}c^2} f^{-4/3} \int d \log_{10} \Mc \int dz
  (1+z)^{-1/3} \Mc^{5/3} \\
\times \frac{d^3N}{dV_{\mathrm{c}}dzd\log_{10}\Mc}\,, \numberthis 
\label{eqn:hsquared}
\end{align*}
where $z$ is the redshift and $\Mc$ is the chirp mass related to the binary component masses ($m_1$, $m_2$) by $\Mc=(m_1m_2)^{3/5}/(m_1+m_2)^{1/5}$. The integral sums over the sources in $z$ and $\Mc$ weighted by the distribution of the source population, $d^3N/dV_{\mathrm{c}}dzd\log_{10}\Mc$, the number of binary mergers per co-moving volume, redshift and (rest-frame) chirp mass interval. We choose a simple model for this, described by
\begin{align*}
\frac{d^3N}{dV_{\mathrm{c}}dzd\log_{10}\Mc} = {\dot n}_0\, \left[\left( \frac{\Mc}{10^7\Msol}\right)^{-\alpha} \exp^{-(\Mc/\Mstar)}\right]\\ \times 
\left[(1+z)^{\beta} \exp^{-(z/z_0)}\right] \frac{dt_R}{dz}\,, \numberthis 
\label{eqn:modeldNdVdzdlogM}
\end{align*}
where $t_R$ is the time in the source rest-frame (here we use $H_0 =70\,\mathrm{km~s}^{-1}\mathrm{Mpc}^{-1}$, $\Omega_M=0.3$, $\Omega_{\Lambda}=0.7$ and $\Omega_k=0$).  Following general astrophysical assumptions, we consider a scenario where the GW background is produced by MBHBs in the redshift and chirp mass range of $0\leq z\leq 5$ and $10^6\leq\Mc/\Msol\leq10^{11}$. These ranges set the integration limits of Eqn.~(\ref{eqn:hsquared}).

The model is described by five parameters. The parameter ${\dot n}_0$ is the normalised merger rate per unit rest-frame time, co-moving volume and logarithmic (rest-frame) chirp mass interval.  The parameters $\beta$ and $z_0$ describe the distribution of the sources in redshift.  The parameter $\beta$ controls the low-redshift power-law slope and the parameter $z_0$ the high-redshift cut-off for the distribution; the peak of the merger rate $d^2N/d t_R d V_{\mathrm{c}}$, corresponds to a redshift ($z_0 \beta -1$).  The parameters $\alpha$ and $\Mstar$ provide a similar description of the chirp mass distribution. The model was chosen to capture the expected qualitative features of the cosmic MBH merger rate without restricting to any particular merger history; for example, it can reproduce rates extracted from merger tree models \citep{2003ApJ...582..559V,SesanaVecchioColacino:2008}, and large scale cosmological simulations of structure formation \citep{SpringelEtAl_MilleniumSim:2005,SesanaVecchioVolonteri:2009}.

The characteristic amplitude has a simple power-law scaling, and we can re-write Eqn.~(\ref{eqn:hsquared}) as
\begin{equation}
h(f) = A_\mathrm{1yr} \left(\frac{f}{f_\mathrm{1yr}}\right)^{-2/3}\,,
\label{eqn:hc_PTA}
\end{equation}
where $A_\mathrm{1yr}$ is the characteristic amplitude at the reference frequency $f_\mathrm{1yr} = 1\,\mathrm{yr}^{-1}$, which is customarily used when quoting limits in the PTA literature. A single number, the amplitude $A_\mathrm{1yr}$, carries the whole information about the merging history of MBHBs (within the model considered in this paper), that one wishes to reconstruct from the observations.

\begin{figure*}
\includegraphics[width=0.3\textwidth]{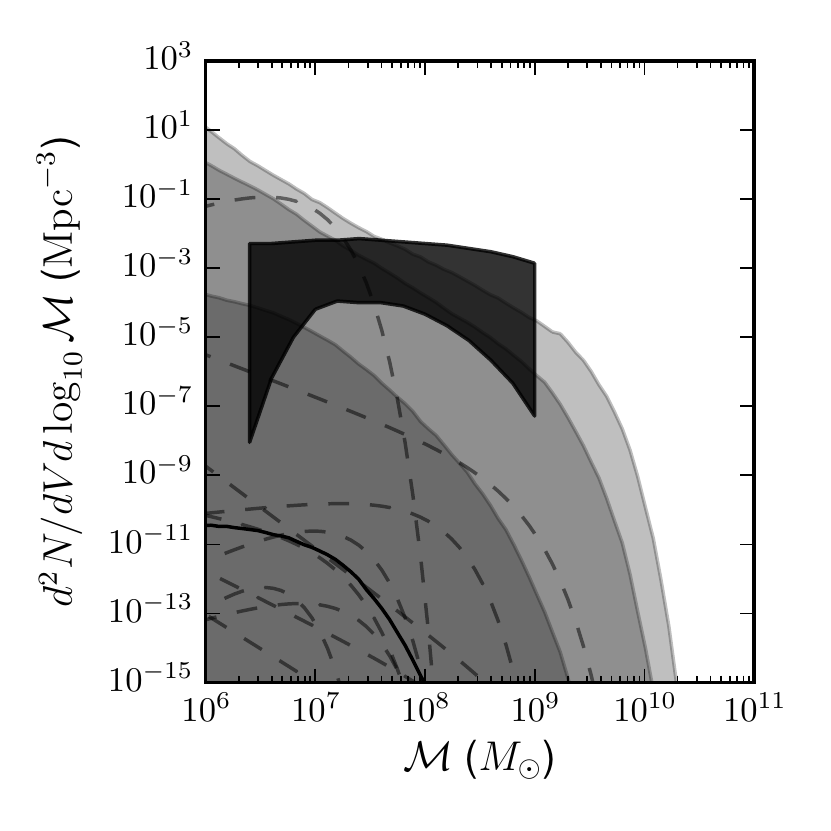}
\includegraphics[width=0.3\textwidth]{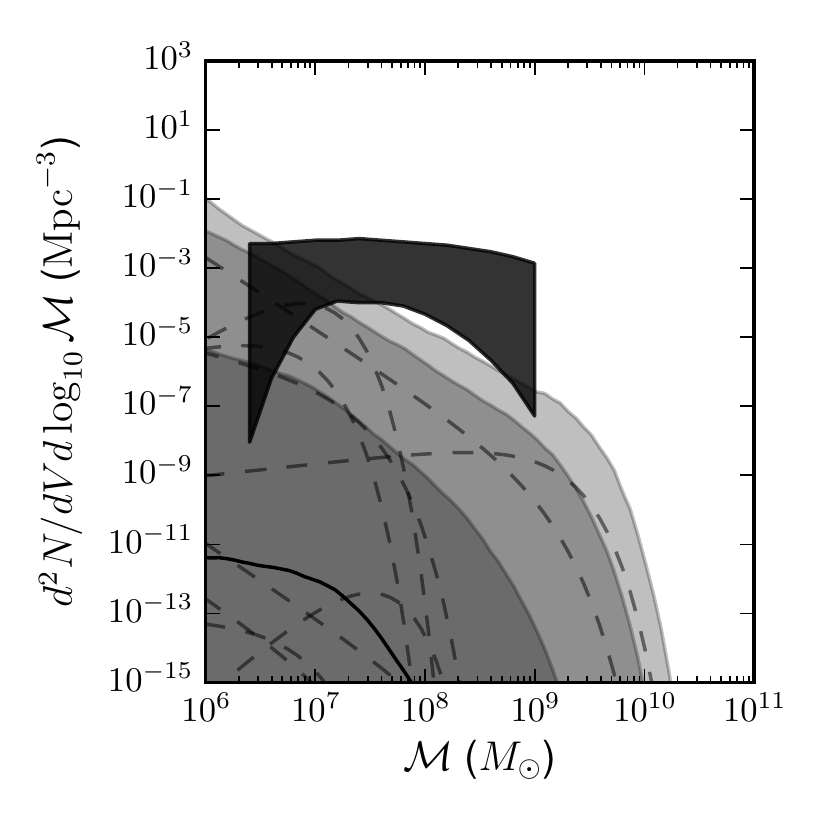}
\includegraphics[width=0.3\textwidth]{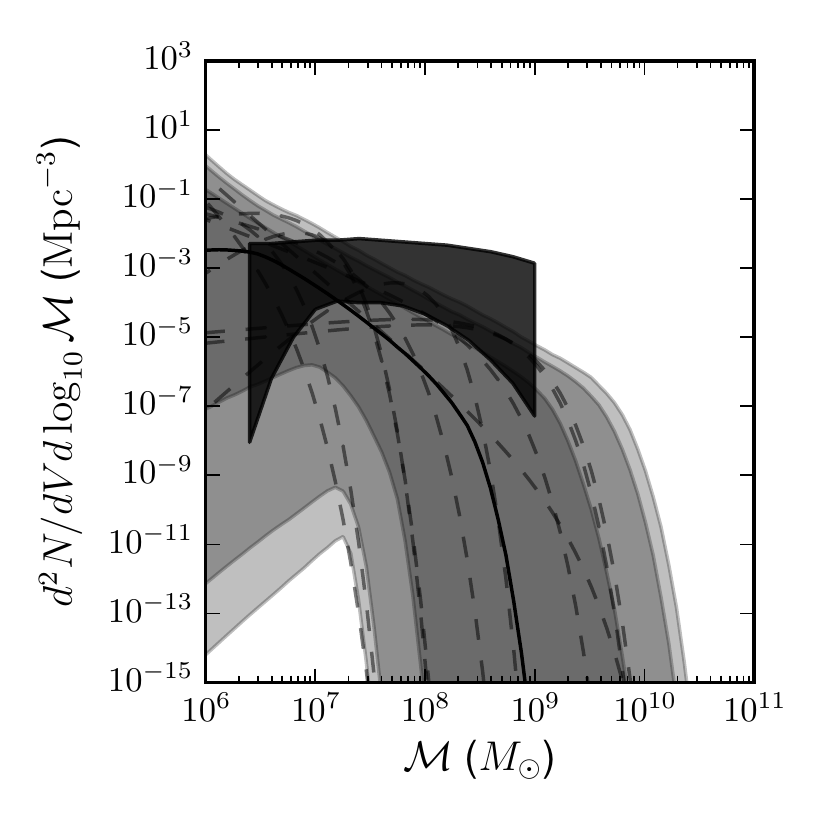}\\
\includegraphics[width=0.3\textwidth]{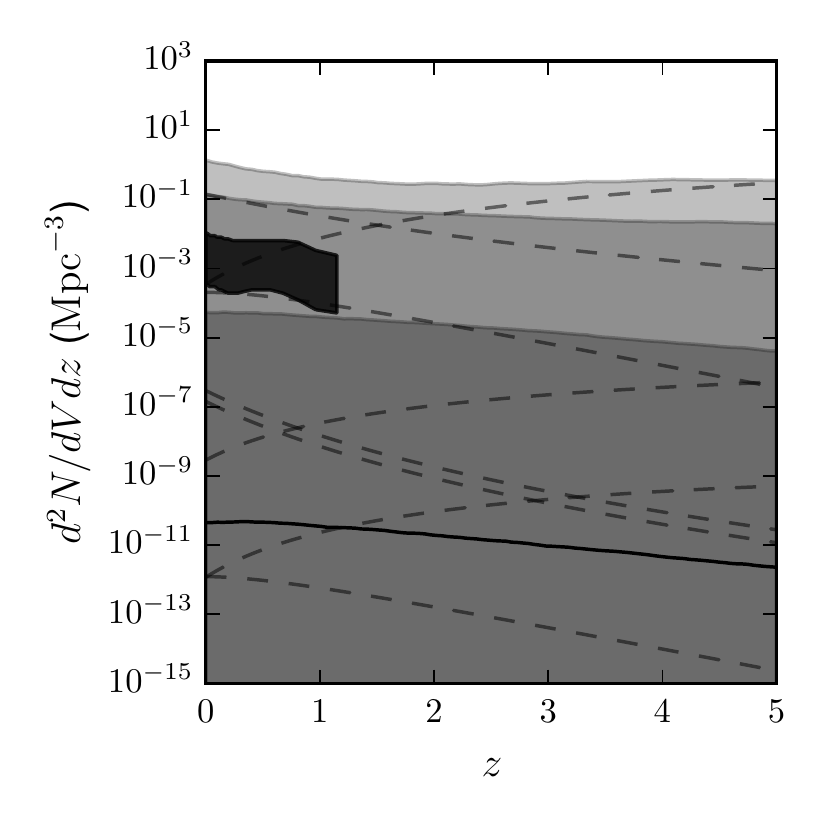}
\includegraphics[width=0.3\textwidth]{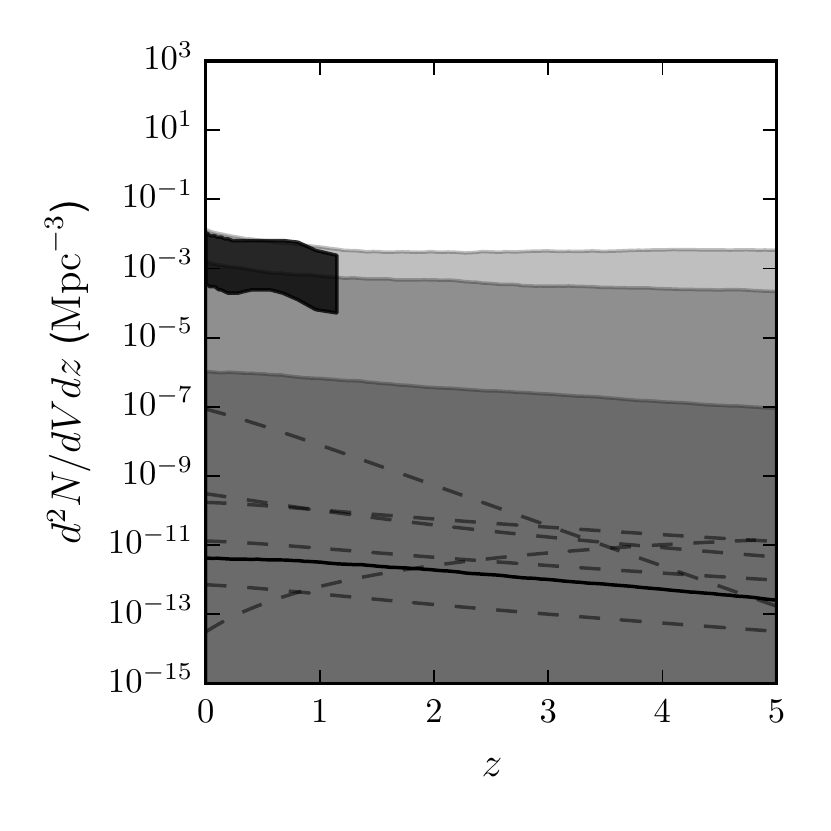}  
\includegraphics[width=0.3\textwidth]{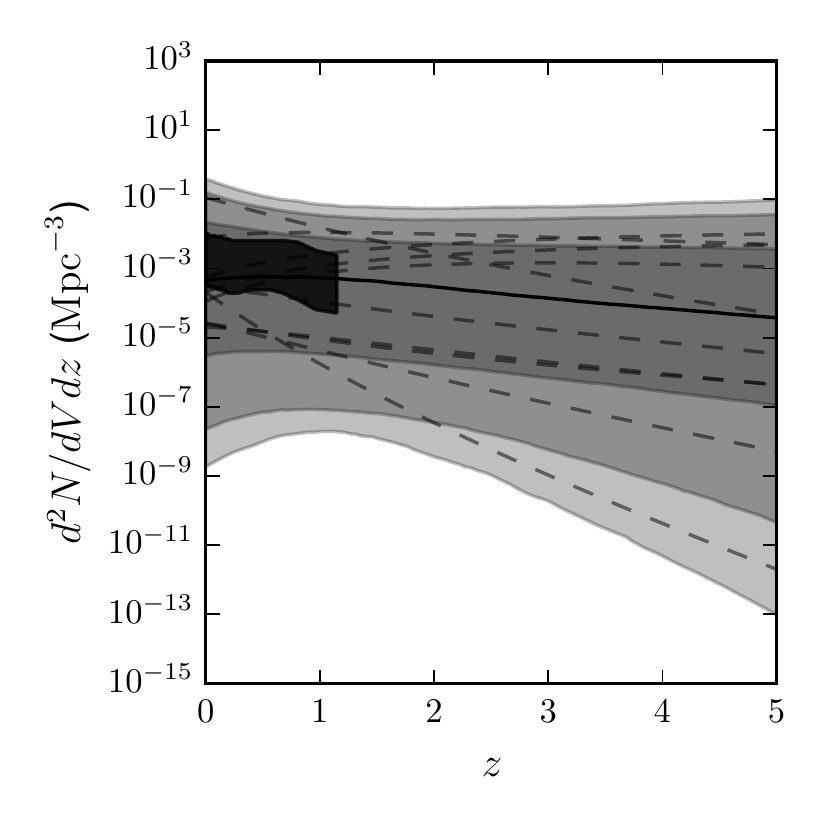}  
\caption{Posteriors for the merger rate density. The top row shows the merger rate density in chirp mass (integrated over redshift), $d^2N/dV_{\mathrm{c}}d\log_{10}{\cal M}$ and the bottom row in redshift (integrated over chirps mass), $d^2N/dV_{\mathrm{c}}dz$, for two 95\% confidence upper limits at $1\times10^{-15}$ (left) and  $1\times10^{-16}$ (centre) and a detection at $1\times 10^{-16}$ (right), as described in the text. We consider contributions to the gravitational wave background from massive black holes in the chirp mass range $10^{6}\le \Mc / M_\odot\le 10^{11}$ and redshift range $0\le z \le 5$. The solid black line gives the posterior median; dark grey, mid-grey and light-grey bands show the central 68\%, 95\%, and 99\% credible interval, respectively. The dashed lines show draws from the posterior. For comparison, the overlaid dark areas represent the 99.7\% confidence regions predicted by the MBH assembly models of \protect\cite{Sesana:2013}. For these models, we show only the chirp mass in the range $\approx 10^6\Msol - 10^9\Msol$ as outside this interval the lower percentile is zero. For the redshift range, these models only consider MBHB mergers for  $z\lesssim1.3$.
}\label{fig:dNdVdM_dNdVdz_M6cut}
\end{figure*}

\subsection{Method}
\label{sec:methods}

The objective is to put constraints on the population parameters, which we denote by $\mathbf{\theta}$, given the results of PTA analyses. In our case $\mathbf{\theta}$ is a 5-dimensional parameter space, $\mathbf{\theta} = \{{\dot n}_0, \beta, z_0, \alpha, \Mstar\}$. We want to compute the posterior density function (PDF) of the parameters given PTA observations denoted by $d$. The population parameters fully specify the gravitational wave signal $h(f; \mathbf{\theta})$ (Eqs.~(\ref{eqn:hsquared}) and~(\ref{eqn:hc_PTA})), which in turn specifies the statistical properties of the GW-induced deviations to pulse arrival times, the PTA observable. Given data from pulsar timing and our model for the merger rate (Eqn~(\ref{eqn:modeldNdVdzdlogM})), we use Bayes' theorem to find the posterior distribution of the model parameters
$p(\theta|d)$,
\begin{equation}
  p(\theta | d) = \frac{p(\theta) p(d | \Ayr(\theta))}{p(d)}\,,
\end{equation}
where $p(d | \Ayr(\theta))$ is the PTA likelihood for a given stochastic background, $h(f; \theta)$, $p(\theta)$ is the prior on the model parameters and $p(d)$ is the evidence.  In standard analysis of the PTA data, constraints are put on the GW characteristic amplitude at periods of one year, $\Ayr$, which in turn is a function of the parameters of the underlying population, specified by $h(f; \mathbf{\theta})$. The PTA analysis uses a likelihood function, $p(d | \Ayr(\theta))$, which we approximate as described below.  Our method does not rely on this approximation; we use it only for analytical convenience in this paper.  If a given PTA analysis provides a posterior distribution for $\Ayr$ then a straightforward re-weighting can produce the corresponding likelihood required for our analysis (if flat priors on $\Ayr$ are used in the analysis then the re-weighting is trivial because the posterior and the likelihood are proportional to each other).

In this paper we consider the two cases in which the PTA analysis provides either an upper-limit or a detection. For the upper-limit scenario we model $p(d|\Ayr)$ using a Fermi-like distribution: 
\begin{equation}
p_\mathrm{ul}(d|\Ayr) \propto \left(\exp((\Ayr-A_\mathrm{ul})/\sigma_\mathrm{ul})+1\right)^{-1},
\label{eq:pul}
\end{equation}
where $A_\mathrm{ul}$ is the upper-limit value returned by the actual analysis and the sharpness of the tail-off, $\sigma_\mathrm{ul}$ can be adjusted to give an upper limit with a chosen confidence, which we set at 95\%. 
We model a detection scenario using a Gaussian in the logarithm of $\Ayr$: 
\begin{multline}
p_\mathrm{det}\left(d|\Ayr \right) \propto \exp(-(\log_{10}(\Ayr)-\log_{10}(A_\mathrm{det}))^2/2\sigma_\mathrm{det}^2)
\label{eq:pdet}
\end{multline}
at a chosen level of detection, $A_\mathrm{det}$. We choose the width of the detection to be $\sigma_{\mathrm{det}} = 0.2$. We compute the marginalised distribution on the model parameters $\theta$ using two independent sampling techniques, to verify the results of our analysis: a nested sampling approach~\citep{VeitchVecchio:2010} and {\it emcee}, an ensemble Markov Chain Monte Carlo sampler \citep{ForemanMackayEtAl_emcee:2013}. 

Our priors on the model parameters are set as follows.  We use a prior on ${\dot n}_0$ that is flat in $\log_{10} {\dot n}_0$ down to a lower limit, which we set to ${\dot n}_0 = 10^{-20} \mathrm{Mpc}^{-3} \mathrm{Gyr}^{-1}$, after which it is flat in ${\dot n}_0$ to zero. The prior upper-bound is set to $10^{3} \mathrm{Mpc}^{-3} \mathrm{Gyr}^{-1}$. This value is set by the ultra-conservative assumption that all the matter in the Universe is formed by MBHs. Our prior allows for the number of mergers to span many orders of magnitude (flat in $\log$) but avoids divergence as ${\dot n}_0 \to 0$. It also allows for the {\it absence} of MBHB binaries merging within an Hubble time. The priors for the other parameters are uniform within ranges that incorporate values that give a good fit to semi-analytical merger tree models \citep[see e.g.][]{SesanaVecchioColacino:2008, SesanaVecchioVolonteri:2009, Sesana:2013}: $\alpha\in [-3.0, 3.0]$, $\beta\in [-2.0, 7.0]$, $z_0\in [0.2, 5.0]$ and $\log_{10}\Mstar/M_\odot \in [6.0, 9.0]$.
While our prior allows for parameter values that can reproduce the merger rates of detailed models, it is uninformative in that we do not assume that the merger rate distribution {\it must} take values from those models. Our priors reflect large theoretical uncertainties about MBHB formation and evolution scenarios, and the lack of any confirmed MBHB candidate,~\citep[see however][]{GrahamEtAl2015_PossibleSMBHBinary,LiuEtAl_SMBHB_redshift2:2015,GrahamEtAl_CatalinaSurvey:2015}.

Our method is summarised as: (i) produce a likelihood for $\Ayr$ (in the case of an actual analysis by using smoothed posterior samples from PTA results, re-weighted if necessary depending on the prior), (ii) choose a model for the merger rate of MBHBs, (iii) produce posterior density functions for the model parameters from which we can infer properties of the MBHB population.

\section{Results}
\label{sec:Results}

Current upper-limits on the GW stochastic background obtained recently are $\Ayr^{(95\%)} = 1\times 10^{-15}$, $1.5\times 10^{-15}$, $3\times 10^{-15}$ for the PPTA~\citep{ShannonEtAl_PPTA:2015}, NANOGrav~\citep{NANOGrav9year:2015}, and the EPTA~\citep{LentatiEtAl_EPTA:2015} respectively. The sensitivity gain provided by the addition of new pulsars to the PTAs and more recent data sets may allow in the short-to-mid term to reach a sensitivity below $\Ayr = 1.0\times 10^{-15}$, and in the more distant future $\Ayr \sim 10^{-16}$~\citep{SiemensEtAl:2013, RaviEtAl:2015}. As a consequence, here we consider three PTA analysis outcomes: (i) an upper-limit at 95\% confidence of $1\times 10^{-15}$, which represents the present state of play and either (ii) an upper-limit (at 95\% confidence) of $1\times 10^{-16}$ or (iii) a detection at the same level, that is $A_\mathrm{det} = 1\times 10^{-16}$, and $\sigma_\mathrm{det} = 0.2$ in Eqn.~(\ref{eq:pdet}), which describes possible results coming from the expected improvements of the PTA sensitivity in the next five-to-ten years. 

\begin{figure}
  \begin{center}
  \includegraphics[width=.48\textwidth]{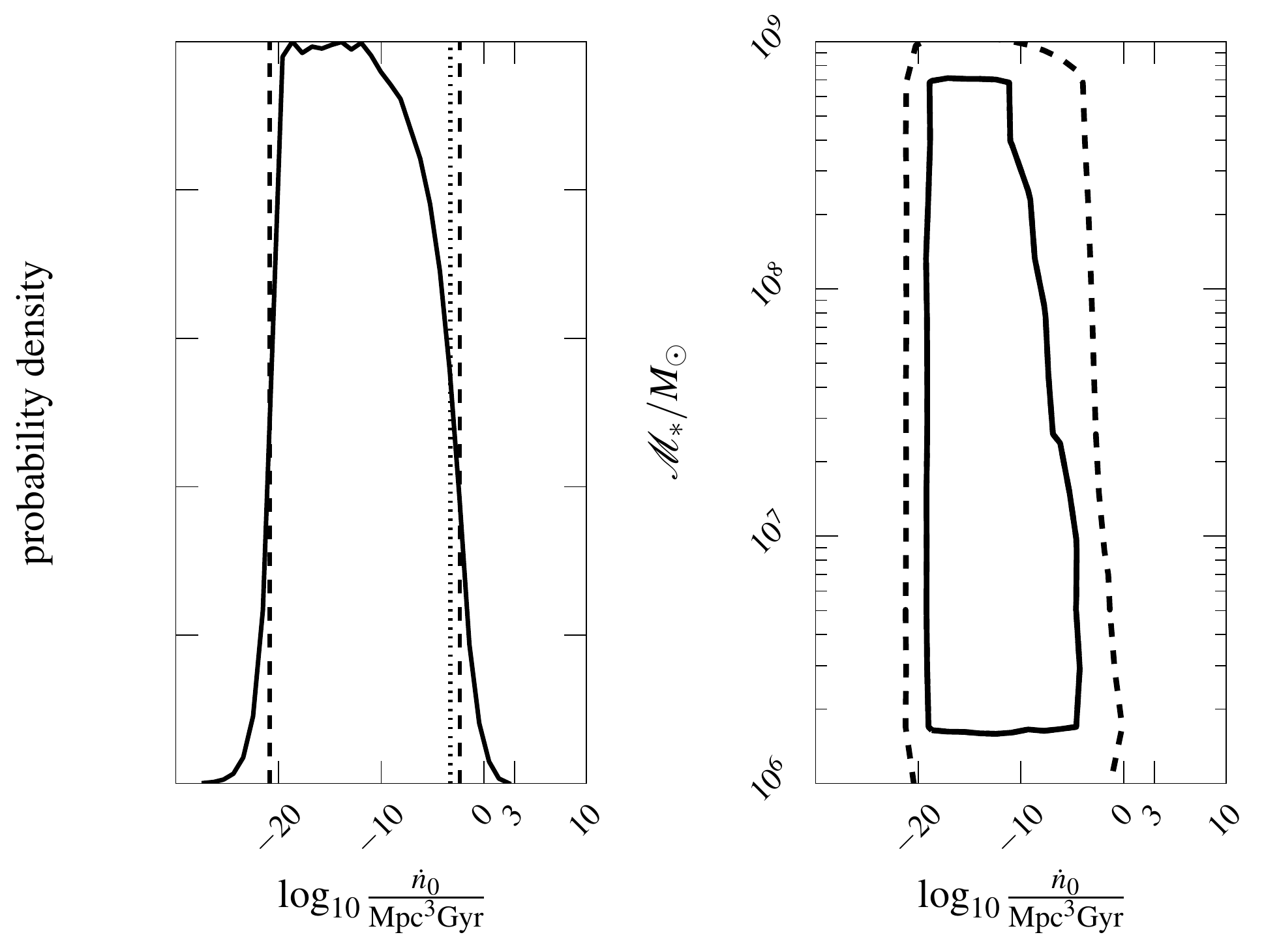}
  \end{center}
\caption{Marginalised posterior distributions for selected astrophysical parameters for the case of 95\% upper-limit of $1\times 10^{-15}$, which corresponds to the current status of the observations. The marginalised PDF on the merger rate parameter, $\dot{n}_0$  is shown in the left panel, and the marginalised PDF on $(\Mstar, \dot{n}_0)$ in the right panel, where the contours mark the 67\% (solid) and 95\% (dashed) confidence regions. In the left panel, the dashed lines mark the 95\% confidence width ($-20.8 \le \log [\dot{n}_0/\mathrm{Mpc}^{-3}\, \mathrm{Gyr}^{-1}]\le -2.3$) while the dotted line marks the 95\% upper limit ($\log [\dot{n}_0/\mathrm{Mpc}^{-3}\, \mathrm{Gyr}^{-1}]\le -3.3$). The left hand side of the distribution in $\log (\dot{n}_0/\mathrm{Mpc}^{-3}\, \mathrm{Gyr}^{-1})$ follows our prior, while the right hand side is determined by the PTA upper limit.}
\label{fig:posterior_results_1e-15}
\end{figure}

\begin{figure}
  \begin{center}
  \includegraphics[width=.48\textwidth]{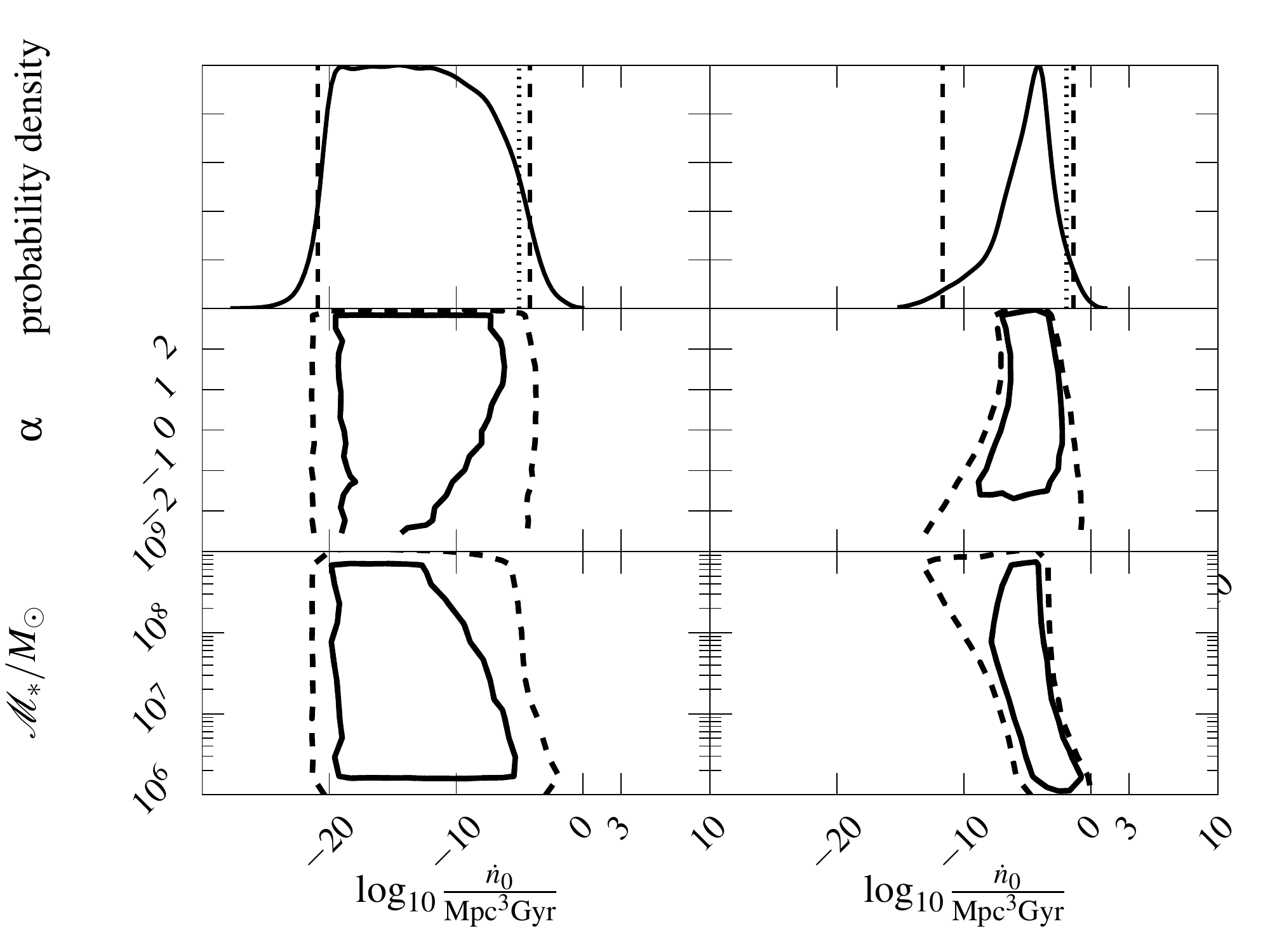}
  \end{center}
  \caption{Posterior distribution for the upper limit (left) and detection (right) at $1\times 10^{-16}$. The top panels show the one dimensional posterior distribution for the merger rate parameter, $\dot{n}_0$. The dashed lines mark the 95\% confidence width (upper limit: $-20.9 \le \log_{10} (\dot{n}_0/\mathrm{Mpc}^{-3}\, \mathrm{Gyr}^{-1})\le -4.2$; detection: $-11.7 \le \log_{10} (\dot{n}_0/\mathrm{Mpc}^{-3}\, \mathrm{Gyr}^{-1})\le -1.3$) and the dotted line the 95\% upper limit (upper limit: $\log_{10} (\dot{n}_0/\mathrm{Mpc}^{-3}\, \mathrm{Gyr}^{-1}) = -5.0$; detection: $ \log_{10} (\dot{n}_0/\mathrm{Mpc}^{-3}\, \mathrm{Gyr}^{-1}) = -1.9$). The central and bottom panels show the two dimensional posterior distributions for ${\dot n}_0$ with the mass parameters $\alpha$ and $\Mstar$. The solid and dashed contours mark the 67\% and 95\% confidence regions respectively.
\label{fig:posterior_results_1e-16}  }
\end{figure}

The main results of our analysis are summarised in Fig.~\ref{fig:dNdVdM_dNdVdz_M6cut}, which shows the inferred posterior distribution of the merger history of MBHBs in terms of the MBHBs co-moving volume merger density per redshift and (logarithm of) chirp mass intervals. 
Fig.~\ref{fig:posterior_results_1e-15} provides PDFs on selected parameters based on current PTA limits, and Fig.~\ref{fig:posterior_results_1e-16} provides a similar summary for a future limit or a detection at the level described above.

We consider first the implications of current limits. The PDFs on the parameters $\dot{n}_0$ and $\Mstar$ of the model are shown in Fig.~\ref{fig:posterior_results_1e-15}; we do not provide the equivalent plots for $\alpha$, $\beta$ and $z_0$ as they are equivalent to the prior. Fig.~\ref{fig:posterior_results_1e-15} clearly shows that the present PTA limits enable us to reduce the allowed normalization of the MBHB merger rate density to $\dot{n}_0 \lesssim 5\times 10^{-3} \mathrm{Mpc}^{-3} \mathrm{Gyr}^{-1}$ with 95\% confidence, but yield no additional constraints on the other parameters of the model.

Our model contains parameters describing the shape of the merger rate distribution in redshift and chirp mass. The PDFs of those parameters induce a posterior density on $d^2N/dV_{\mathrm{c}}d\log_{10}{\cal M}$ and $d^2N/dV_{\mathrm{c}}dz$, integrating over redshift and chirp mass respectively, shown in Fig.~\ref{fig:dNdVdM_dNdVdz_M6cut}. We see that current observations limit the maximum merger density as a function of mass, but place no constraints on the shape of the distribution.  
The corresponding number of sources per frequency bin that contribute to the signal is $ dN/df d\log_{10}\mathcal{M} \propto f^{-11/3}$, and we find that for masses above a few $ \times 10^9 M_\odot$, our upper limit on $d^2 N / dV_{\mathrm{c}}d\log_{10}{\cal M}$ implies that at a frequency around $1.8$ nHz there is fewer than one source per frequency bin (taken to be $\Delta f=1/T$, with $T=17.66$ yr, the timespan of current EPTA datasets~\citep{LentatiEtAl_EPTA:2015}). 
This means that at those large masses, the assumption that the observed GW signal is stochastic is violated, and our analysis cannot be used to constrain the exact shape of the mass function here (in this case a different PTA search approach would be necessary, see e.g.~\cite{BabakEtAl:2015,ArzoumanianEtAl:2014,TaylorEtAl_EPTAanisotropyAnalysis:2015}). While current PTA observations provide feeble constraints on the shape of the mass distribution, they yield no information about the redshift distribution. The bottom-left panel of Fig.~\ref{fig:dNdVdM_dNdVdz_M6cut} shows no structure in $d^2N/dV_{\mathrm{c}}dz$.

It is useful to compare these results to limits on the MBHB merger rates implied by binary candidates reported in the literature and to specific theoretical models. Let us first consider what is known {\it observationally} today. A few MBHB candidates have been reported recently. \cite{GrahamEtAl2015_PossibleSMBHBinary} suggested the possible observation of a MBHB at redshift $z=0.2784$ with (rest frame) total mass $\log(M/\Msol)\sim8.5$ and period of $\sim 1884$ days. \cite{LiuEtAl_SMBHB_redshift2:2015} reported the observation of a potential MBHB at $z=2.060$ with a shorter period of 542 days and primary MBHB mass $\log(M/\Msol)\sim 9.97$. Using the redshift to calculate the enclosed volume and the binary parameters for the time to merger we can estimate the predicted rate from each of these observations. Assuming that these two systems are indeed MBHBs, and that their constituents are of comparable mass, they imply merger rates of $\approx 3\times10^{-7}\, \mathrm{Mpc}^{-3} \mathrm{Gyr}^{-1}$ and $\approx 0.1\, \mathrm{Mpc}^{-3} \mathrm{Gyr}^{-1}$ (this latter number takes into account that the source has been found in an analysis of only one of the Pan-STARRS1 Medium Deep Survey fields of 8 deg$^{2}$). In turn they yield a merger density  $d^2N/dV_{\mathrm{c}}d\log_{10}{\cal M}$ $\approx 10^{-6}\, \mathrm{Mpc}^{-3}$ at ${\cal M}\approx 3\times 10^{8}\,\Msol$ and $\approx 1\, \mathrm{Mpc}^{-3}$ at ${\cal M}\approx 10^{10}\,\Msol$, respectively.  The upper left panel of Figure~\ref{fig:dNdVdM_dNdVdz_M6cut} clearly shows that the rate density inferred from \cite{GrahamEtAl2015_PossibleSMBHBinary} is consistent with current upper limits, while that inferred from \citet{LiuEtAl_SMBHB_redshift2:2015} is several orders of magnitude above the 99\% credible interval implied by current PTA results. It is therefore unlikely that this source is a MBHB with the claimed parameters.  Other proposed MBHBs in the literature \citep{Valtonen2012,Kun2014} imply merger density estimates of $5 \times 10^{-5}\, \mathrm{Mpc}^{-3}$ at $\Mc \approx 10^9 M_\odot$ and $3 \times 10^{-6} \, \mathrm{Mpc}^{-3}$ at $\Mc \approx 3.5 \times 10^{8} \, M_\odot$, which are consistent with the current upper limits.

On the theoretical side, current limits are consistent with the assumption that most Milky-Way-like galaxies contain a MBH in the mass range considered here that undergoes $\sim 1$ major merger in an Hubble time. The density of Milky-Way-like galaxies is $10^{-2}\,\mathrm{Mpc}^{-3}$, which yields an estimate of $d^2N/dV_{\mathrm{c}}d\log_{10}{\cal M} \sim 10^{-3}\,\mathrm{Mpc}^{-3}$, which is consistent with our results at $\Mc\sim 10^6\Msol$, appropriate for a typical MBHB forming in the merger of Milky-Way-like galaxies. We also compare the limits on the $d^2N/dV_{\mathrm{c}}d\log_{10}{\cal M}$ and $d^2N/dV_{\mathrm{c}}dz$ with specific distributions obtained from predictions of astrophysical models for the cosmic assembly of MBHs. We consider the models presented in \cite{Sesana:2013}, extended to include the most recent MBH-galaxy scaling relations \citep{2013ARA&A..51..511K}. These models produce a central 99\% interval of $\Ayr \in [2\times10^{-16},4\times10^{-15}]$. The 99.7\% confidence region in the merger density from those models is marked by a dark-shaded area in each panel of Fig.~\ref{fig:dNdVdM_dNdVdz_M6cut}. Two conclusions can be drawn: (i) present MBHB population models are consistent with current PTA limits; (ii) those models are drawn from a very restricted prior range of the parameters that control the evolution of MBHBs, driven by specific assumptions on their assembly history. For example, in those models there is a one-to-one correspondence between galaxy and MBH mergers. Our results are consistent with the conclusions drawn by~\cite{ShannonEtAl_PPTAgwbg:2013} about the implications of the PPTA limit for the MBHB merger history. However, since \cite{ShannonEtAl_PPTAgwbg:2013} consider specific models that lie close to the upper end of the 99\% credible range allowed by current limits, they emphasise the fact that PTA limits might soon be in tension with those specific classes of models.

We turn now to consider what we could infer about the MBHB merger history in the future as PTA sensitivity increases. For definiteness we consider both an upper limit and a detection at the level of $A_\mathrm{1yr} = 10^{-16}$. Selected marginalised PDFs on the model parameters are shown in Fig.~\ref{fig:posterior_results_1e-16}, where we see a slight correlation in the 2-D marginalised PDF of $(\Mstar, \dot{n}_0)$, as expected. This is simply explained by considering the Schechter-like mass profile of Eqn.~\ref{eqn:modeldNdVdzdlogM}: as the characteristic mass $\Mstar$ decreases, and therefore the exponential cut-off of MBHB progressively depletes the high-mass portion of the population, a given value of the GW characteristic amplitude allows for a larger overall normalisation, $\dot{n}_0$. The posterior MBHB merger densities per logarithm of chirp mass and redshift are shown in Fig.~\ref{fig:dNdVdM_dNdVdz_M6cut}. For the case of an upper-limit the results are qualitatively similar to the case of the present PTA upper-limit, simply scaled accordingly. In particular, despite the much tighter limit on the overall merger rate we are still unable to place any meaningful constraint on the redshift distribution of merging MBHBs. The overall merger density as a function of redshift shifts by two orders of magnitude and the same is true for the merger density as a function of mass. Note that a non-detection at this level might pose a serious challenge to currently favoured theoretical MBH assembly models with simple black hole dynamics, as shown in the upper-centre panel of Fig.~\ref{fig:dNdVdM_dNdVdz_M6cut}.  

In the case of detection the posterior on the shapes of the merger rate distribution in redshift and chirp mass are plotted on the right panels of Fig.~\ref{fig:dNdVdM_dNdVdz_M6cut}. We still obtain essentially no bounds on the shape of the merger rate density in redshift. We also obtain no meaningful lower bound on the merger rate density for chirp masses. That is, there is no chirp mass at which we can bound the merger density above a rate physically indistinguishable from zero; we know that some MBHBs merge, but we cannot determine {\it which ones}. Additional information, such as theoretical assumptions, electromagnetic observations constraining the mass spectrum of merging black holes (like those discussed earlier in this Section), or gravitational wave observations that measure the binary mass spectrum directly (such as those of an eLISA-like instrument \citep{2013arXiv1305.5720C}) , are required to place any constraints on the masses of the merging systems. For example, if we accept the priors provided by \cite{Sesana:2013}, the mass function of merging MBHBs can be determined more precisely, as shown by the overlap between our posterior and the dark band in the upper--right panel of Fig.~\ref{fig:dNdVdM_dNdVdz_M6cut}.

\section{Conclusions}
\label{sec:Conclusions}

We have considered the implications of current PTA limits on the GW stochastic background to constrain the merger history of MBHBs. Using a general model for the mass and redshift evolution of MBHBs in circular orbit driven by radiation reaction, we find that existing PTA results alone place essentially no constraints on the merger history of MBHBs. We also find that even with an increase in amplitude sensitivity of an order of magnitude, and assuming that a detection is made, no bounds can be put on the functional form of the merger rate density in redshift and chirp mass unless additional information coming through a different set of observations is available.

Finally we want to caution the reader that the results presented here apply only within the model assumptions that have been made. We have considered a generic (and well justified) functional form for the MBHB merger rate density, but if one chooses a significantly different form (and associated priors for the parameters), results could be different (even radically). Moreover, it has been suggested that physical effects other than radiation reaction, such as gas and/or interactions with stars~\citep[e.g.][]{2011MNRAS.411.1467K,Sesana:2013CQGra,Sampson:2015}, could affect the evolution of MBHBs. These effects are not included in our model, and their impact on astrophysical inference needs to be evaluated in the future.

\bibliographystyle{mnras}
\bibliography{bibliography_v3}

\label{lastpage}

\end{document}